\documentclass[aps,prl,twocolumn,superscriptaddress,nofootinbib,floatfix]{revtex4-2}
\usepackage[T1]{fontenc}
\usepackage{newtxtext,newtxmath}
\usepackage{amsmath,bm,mathtools}
\usepackage{graphicx,xcolor,microtype,placeins,hyperref}
\usepackage{orcidlink}
\hypersetup{
 colorlinks=true,
 linkcolor=blue!45!black,
 citecolor=blue!45!black,
 urlcolor=blue!45!black,
 pdftitle={A Zoology of Quantum Turing Patterns},
 pdfauthor={Kazuki Ikeda},
 pdfsubject={Different asymptotic noise laws for visible Turing order and Gaussian witnesses of the same ordering mode},
 pdfkeywords={quantum Turing patterns, Lindblad dynamics, Gaussian entanglement, translation symmetry breaking, stripe nematic crossover}}
\graphicspath{{figures/}{../figures/}}
\newcommand{\Diss}{\mathcal D}
\newcommand{\cN}{\mathcal N}
\newcommand{\PhiS}{\Phi_{2}^{(S)}}
\newcommand{\USB}{\mathcal U_{\rm SB}}

\begin{document}
\title{A Zoology of Quantum Turing Patterns}
\author{Kazuki Ikeda\,\orcidlink{0000-0003-3821-2669}}
\email{kazuki.ikeda@umb.edu}
\affiliation{Department of Physics, University of Massachusetts Boston, Boston, MA 02125, USA}
\affiliation{Center for Nuclear Theory, Department of Physics and Astronomy, Stony Brook University, Stony Brook, New York 11794-3800, USA}

\begin{abstract}
We explore quantum Turing pattern zoology, where the same Lindblad equation supports a morphology atlas of stripes, spots, holes, labyrinths, and defects. The stable stripe species provides a quantitatively controlled case in which morphology and Gaussian witness loss separate parametrically. In particular, visible Turing stripes can remain after two Gaussian witness margins associated with the same $k_*$ mode cross zero in a completely positive Lindblad lattice. The witness thresholds on the exact shell fall as $\cN^{-1}$. The stripe nematic threshold tends to a nonzero value at fixed lattice size, time window, and morphology criterion. The ratio of the morphology threshold to either witness threshold therefore grows with $\cN$. Imaging and momentum-resolved covariance measurements probe these sectors separately.
\end{abstract}
\maketitle

\section{Introduction}
Turing's mechanism explains how local dynamics and differential transport select a finite wavelength \cite{Turing1952,CrossHohenberg1993,CrossGreenside2009,GiererMeinhardt1972,MainiPainterChau1997,KondoMiura2010,Krause2021}. Pigment patterns, hair follicle spacing, digit specification, and palatal ridges provide experimental realizations \cite{KondoAsai1995,KondoWatanabeMiyazawa2021,YamaguchiYoshimotoKondo2007,NakamasuTakahashiKanbeKondo2009,SickReinkerTimmerSchlake2006,RaspopovicMarconRussoSharpe2014,EconomouEtAl2012}. Network surveys also find Turing structure without prescribed feedback \cite{PaulAdetunjiHong2024}. A single selected wavelength can support several morphologies. Nearby parameters, preparation, noise, and observation time can yield stripes, spots, holes, labyrinths, defects, or coarsening mixtures \cite{Hoyle2006,Leppanen2004,GuiuSouto2012,Bray1994}.

The stripe and the two witnesses studied below share the wave number $k_*$ and one completely positive Lindblad generator. The generator enters the first moments through $\Theta=(2n_{\rm th}+1)/\cN$ and the canonical covariance through $n_{\rm th}$, producing the two scaling laws derived below.

Turing order has been predicted in cold excitons and observed in a coherent polariton fluid \cite{LevitovSimonsButov2005,Ardizzone2013}. Models with a few activator and inhibitor modes, as well as multimode generators, support quantum Turing instabilities \cite{BandyopadhyayKhatunBanerjee2021,KatoNakao2022,ComparatoGarganoLoFranco2026}. Transverse quantum optical patterns exhibit fluctuation-seeded images, spatial correlations, and squeezing \cite{Gatti1995,Gatti1997,Marzoli1997,Navarrete2010}. Ref.~\cite{Ikeda2026} constructed the Lindblad family used here and obtained stable quantum Turing patterns. Previous studies characterize the instability and its spatial correlations. Whether morphology and shell witnesses fail at the same physical noise remains unresolved.

Gaussian witness crossings remain at bath occupations of order unity, and their thresholds in physical noise therefore fall as $\cN^{-1}$. The stripe nematic threshold instead tends to a nonzero value at fixed lattice size, time window, and morphology criterion. Longer trajectories, the orthogonal stripe, and modest variations of the morphology cutoff leave this ordering unchanged. A multimode covariance sector remains nonclassical beyond the two pairwise shell crossings.

The interval in which stripes remain resolved after the two shell margins cross zero widens with $\cN$. This growing interval contains resolved stripes after both tested covariance margins have crossed zero. Similar separation may arise in driven photonic arrays, polariton fluids, and other bosonic lattices whose first moments carry order while second moments carry the nonclassicality tested here.

\begin{figure*}[t]
 \includegraphics[width=\textwidth]{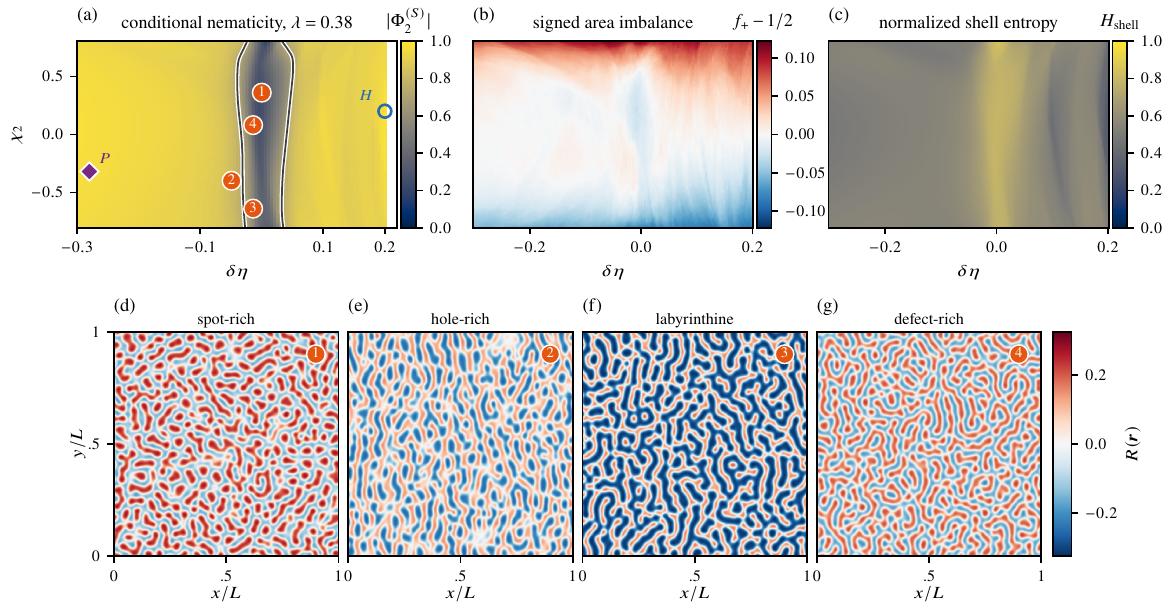}
 \caption{\textbf{Morphology atlas and fields from different preparations.} Conditional stripe nematicity at $\lambda=0.38$ is mapped in panel (a). Gray cells fail the resolved-field criterion and the contour marks $|\Phi_2^{(S)}|=0.72$. The filled diamond P and open circle H denote the two stable stripe branches. Markers 1 through 4 identify the spot-rich, hole-rich, labyrinthine, and defect-rich fields in panels (d) through (g). Area imbalance and shell entropy appear in panels (b) and (c). Every atlas cell starts from the same state. Preparation dependence is tested in the Supplemental Material. The letter P labels the principal stripe branch, while $w_P$ denotes the Glauber--Sudarshan witness.}
 \label{fig:morph}
\end{figure*}

\section{Lindblad model and noise scaling}
We consider a periodic square array of bosonic modes,
\begin{align}
 \dot\rho&=-i[H,\rho]+\sum_\alpha\Diss[L_\alpha]\rho,\nonumber\\
 \Diss[L]\rho&=L\rho L^\dagger-\tfrac12\{L^\dagger L,\rho\}.
 \label{eq:lindblad}
\end{align}
Here $\rho$ is the density operator, $H$ is the Hamiltonian, and $L_\alpha$ are jump operators indexed by the dissipative channel $\alpha$. The generator describes parametrically driven nonlinear resonators with one-photon gain and loss together with two-photon loss. Dissipative and parametric bonds give unequal transport to the two quadratures, nonlinear loss saturates the finite-wave-number instability, and the cubic Hamiltonian breaks parity. One completely positive model fixes both drift and diffusion. Operator coefficients are specified in the Supplemental Material.

The operator $a_{\bm r}$ annihilates a boson at lattice site $\bm r$. Writing
\begin{equation}
 \langle a_{\bm r}\rangle=\sqrt{\cN}\,(q_{\bm r}+ip_{\bm r})/\sqrt2,
 \label{eq:mean_scaling}
\end{equation}
the leading drift is
\begin{align}
 \dot q={}&q+\omega p+D_{qx}\Delta_xq+D_{qy}\Delta_yq\nonumber\\
 &+\chi_2(q^2-p^2)-\nu(q^2+p^2)q,\nonumber\\
 \dot p={}&-\omega q-3p+D_{px}\Delta_xp+D_{py}\Delta_yp\nonumber\\
 &-2\chi_2qp-\nu(q^2+p^2)p.
 \label{eq:qp}
\end{align}
The operators $\Delta_x$ and $\Delta_y$ are periodic nearest-neighbor lattice Laplacians. We use $\nu=4$. The parameter $\cN$ controls local inverse noise and is independent of the number of sites. We set $\omega^2=2\sqrt3-\lambda$ and write the transport anisotropy as $\delta\eta=\eta_{\rm iso}-1$. Reversing the sign of $\delta\eta$ exchanges the preferred stripe direction. The principal branch selects $k_*=\pi/6$ and period 12. The Supplemental Material gives the discrete stability wedge and the orthogonal stable branch.

The physical noise of the first moments is
\begin{equation}
 \Theta=\frac{2n_{\rm th}+1}{\cN},
 \qquad \Theta\ge\cN^{-1}.
 \label{eq:theta}
\end{equation}
The parameter $n_{\rm th}\ge0$ is the thermal bath occupation, and the one-photon rate is $\kappa=2$. Its thermal diffusion coefficient is $\kappa(n_{\rm th}+1/2)/\cN=(\kappa/2)\Theta$. Let $V$ be the symmetrized covariance matrix of canonical quadrature fluctuations about the stripe, and let $J$ be the corresponding linearized drift matrix. These fluctuations obey
\begin{equation}
 \dot V=JV+VJ^{\mathsf T}+D_0+n_{\rm th}D_T,
 \label{eq:noise_sectors}
\end{equation}
where $D_0$ is the diffusion matrix at zero temperature and $D_T$ is the thermal increment. Let $n_{G,k_*}^{c,(0)}$ denote the leading Gaussian crossing before covariance feedback. The label $G=\mathrm{NPT}$ refers to the negative partial transpose witness, while $G=P$ refers to the second-moment Glauber--Sudarshan $P$ witness. Their margins are defined in Eq.~\eqref{eq:witnesses}. Smooth covariance feedback gives
\begin{equation}
 \Theta_{G,k_*}^c(\cN)=\frac{2n_{G,k_*}^{c,(0)}+1}{\cN}+O(\cN^{-2}).
 \label{eq:gaussian_scale}
\end{equation}
The subscript $M$ denotes morphology, and the superscript $c$ labels the stated crossing criterion. At fixed lattice size, time window, and morphology criterion,
\begin{equation}
 \Theta_M^c(\cN)\longrightarrow\Theta_M^{\rm lim}>0,
 \qquad
 \frac{\Theta_M^c(\cN)}{\Theta_{G,k_*}^c(\cN)}\propto\cN.
 \label{eq:separation_law}
\end{equation}

\begin{figure*}[t]
 \includegraphics[width=\textwidth]{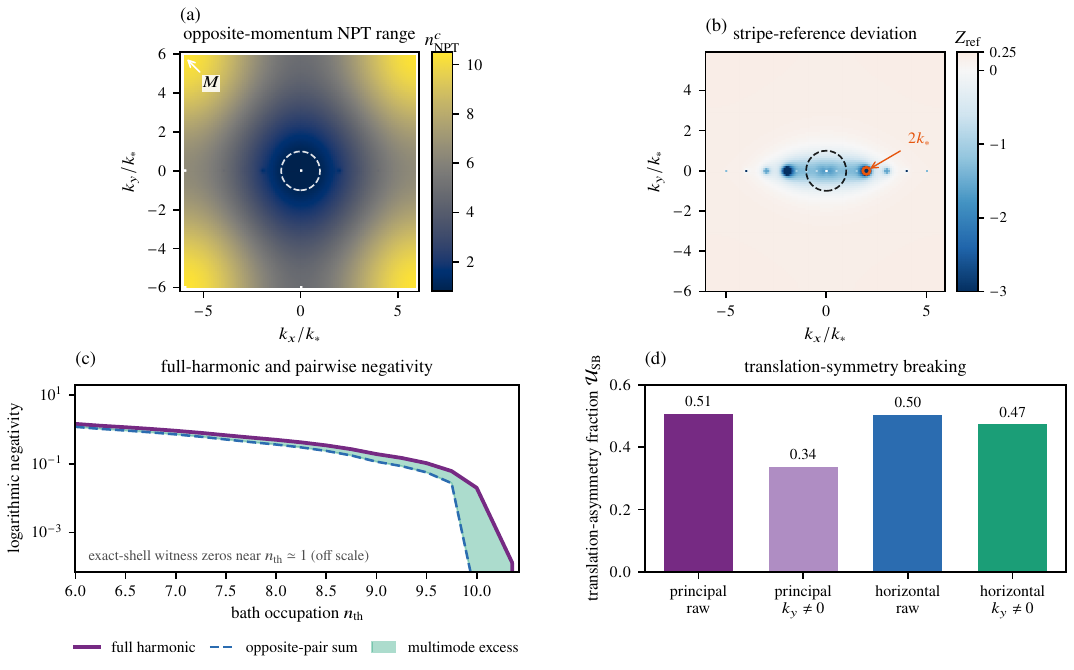}
 \caption{\textbf{Momentum-space Gaussian structure and translation breaking.} Panel (a) maps the noise range certified by the negative partial transpose (NPT) criterion for opposite-momentum pairs. The dashed circle identifies the exact $k_*$ shell, and $M$ is the endpoint invariant under momentum inversion. Panel (b) gives the standardized crossing difference $Z_{\rm ref}$ relative to uniform comparison states made stationary by constant drives. Negative values place the stripe crossing below the family mean, with the contrast concentrated near $2k_*$. The linear scale spans $-3$ to $0.25$, maps zero to white, and clips values below $-3$ to the lower endpoint. Panel (c) shows the scanned Bloch partition with the largest crossing. Its full-sector negativity remains nonzero after the two shell witness crossings, and the shaded area is the full-sector result minus the sum of the tested opposite-momentum reductions. Panel (d) measures translation asymmetry before and after removal of the $k_y=0$ sector containing the nearly neutral translation mode.}
 \label{fig:global}
\end{figure*}

\section{Nonlinear morphology}
For $R=q-\beta p$, with $\beta=(\sqrt3-1)/\sqrt{2\sqrt3}$, let $S(\bm k)=|R_{\bm k}|^2$ and $\theta_{\bm k}=\arg(k_x+ik_y)$. Angular organization on an annulus $\mathcal S$ about $k_*$ is
\begin{equation}
 \Phi_m^{(S)}=
 \frac{\sum_{\bm k\in\mathcal S}S(\bm k)e^{im\theta_{\bm k}}}
 {\sum_{\bm k\in\mathcal S}S(\bm k)}.
 \label{eq:phi}
\end{equation}
The static atlas displays $|\Phi_2^{(S)}|$ only for fields that pass the criteria for real-space contrast and concentration on the shell. Thermal trajectories use the ensemble mean of $|\Phi_2^{(S)}|$ averaged over time. Contrast and Bragg concentration are tracked separately from stripe nematicity.

The four fields in Fig.~\ref{fig:morph} arise from separate preparations at the marked parameter values. The covariance calculation uses the two stationary stripe branches. The Supplemental Material compares eight preparations and variations of timestep, duration, and lattice size.

\section{Momentum-space correlations and translation breaking}
A stripe registry is the spatial phase of its period-12 profile. Linearization at fixed registry gives the canonical covariance. For a tested opposite-momentum pair with covariance $V_{\rm pair}$, define
\begin{equation}
 w_{\rm NPT}=\frac12-\widetilde\nu_-^{\rm PT},
 \qquad w_P=-\lambda_{\min}(V_{\rm pair}-I_4/2).
 \label{eq:witnesses}
\end{equation}
Here PT denotes partial transposition, $\widetilde\nu_-^{\rm PT}$ is the smallest symplectic eigenvalue of the partially transposed pair covariance, $I_4$ is the identity on the four-dimensional pair covariance space, and $\lambda_{\min}$ denotes the smallest ordinary eigenvalue. A positive $w_{\rm NPT}$ certifies Gaussian entanglement for the tested bipartition. A positive $w_P$ detects second-moment nonclassicality in the Glauber--Sudarshan $P$ sense. On the exact $k_*$ shell,
\begin{equation}
 n_{{\rm NPT},k_*}^{c}\simeq1.02,
 \qquad n_{P,k_*}^{c}\simeq1.04.
 \label{eq:shellBounds}
\end{equation}

Most of the broad profile over opposite momentum pairs is reproduced by the uniform comparison states. Stripe order adds a localized contrast near $2k_*$. In the sector identified in Fig.~\ref{fig:global}(c), the negativity of the full covariance remains after the pair witnesses on the exact $k_*$ shell cross zero. The last pairwise boundary across the Brillouin zone approaches the $M$ point, which is invariant under momentum inversion.

Let $I_{\rm can}$ be the identity on the full canonical covariance space, set $C=V-I_{\rm can}/2$, and let $\mathcal T_{\rm fluc}$ average the fluctuation covariance over the 12 translated registries. We quantify translation asymmetry by
\begin{equation}
 \USB=\frac{\|C-\mathcal T_{\rm fluc}(C)\|_F^2}{\|C\|_F^2}.
 \label{eq:USB}
\end{equation}
Here $\|\cdot\|_F$ is the Frobenius norm. After removal of the $k_y=0$ sector containing the nearly neutral translation mode, the zero-bath values are $0.34$ for the principal stripe and $0.47$ for the horizontal stripe. The uniform comparison states are translation invariant.

\begin{figure*}[t]
 \includegraphics[width=\textwidth]{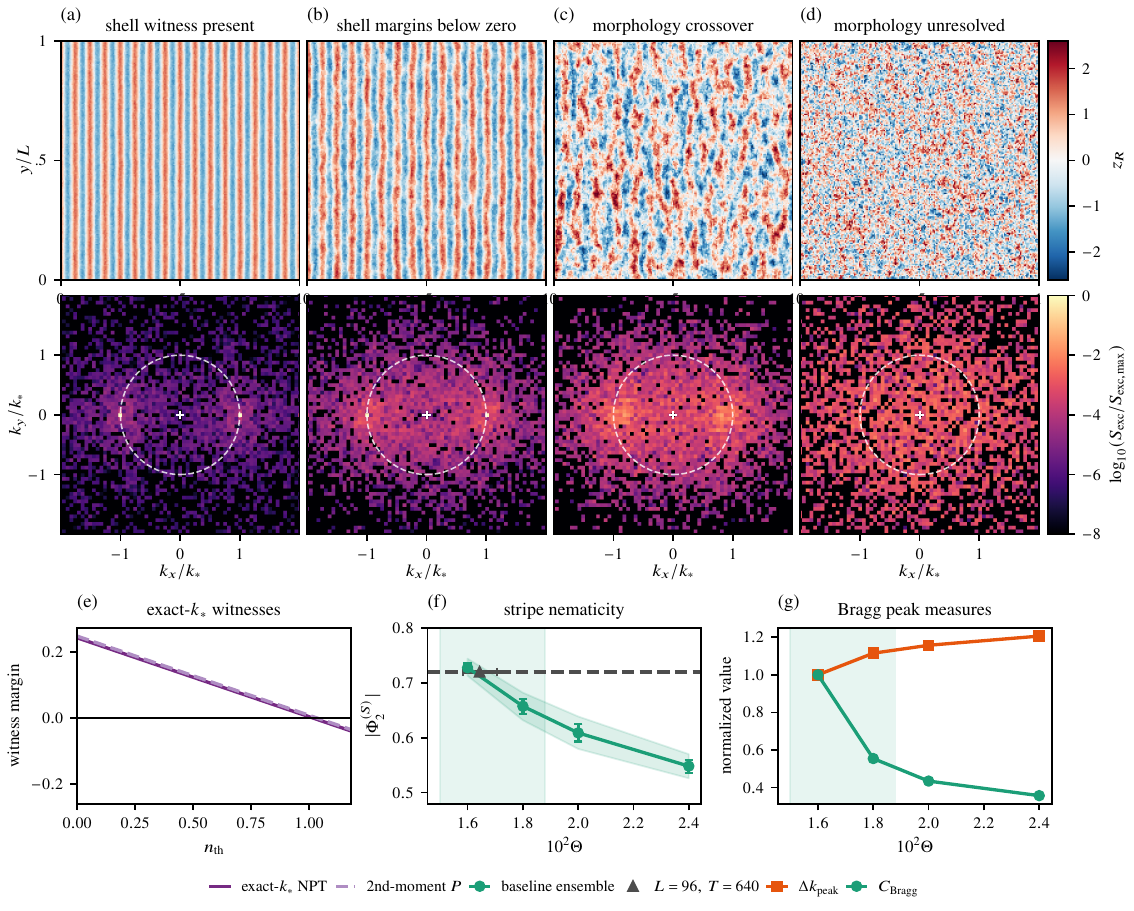}
 \caption{\textbf{Gaussian witness loss and stripe nematic crossover.} Each field in panels (a) through (d) is centered and standardized separately. The panels show a stripe with a positive witness margin on the exact $k_*$ shell, a resolved stripe after both margins on that shell cross zero, the morphology crossover, and an unresolved morphology. Color therefore encodes form on a common dimensionless scale. The crossover estimator uses the trajectory ensemble, and the four snapshots are illustrative. The Supplement identifies the source arrays and retained metadata for these snapshots. The middle row gives the corresponding Fourier powers after background subtraction on common momentum axes and a common logarithmic scale. Witness margins, stripe nematicity, and Bragg peak width and concentration appear in panel (g). The filled triangle in panel (g) gives the $L=96$, $T=640$ estimate.}
 \label{fig:melting}
\end{figure*}

A weak period-12 field fixes the stripe registry without changing the sign of the localized $2k_*$ response. Including covariance feedback shifts the exact-shell crossings by less than $0.3\%$ and leaves the stripe crossing below all uniform comparison states whose linearized drift eigenvalues have negative real parts. Further controls are given in the Supplemental Material.

\section{Separated noise scales}
We locate the stripe nematic crossover where the ensemble mean satisfies $\langle|\Phi_2^{(S)}|\rangle=0.72$. We repeat the calculation for longer trajectories, larger values of $\cN$, the orthogonal stripe branch, and cutoffs from $0.68$ to $0.76$. Real-space contrast and Bragg concentration are monitored separately.

\begin{figure}[t!]
 \includegraphics[width=\columnwidth]{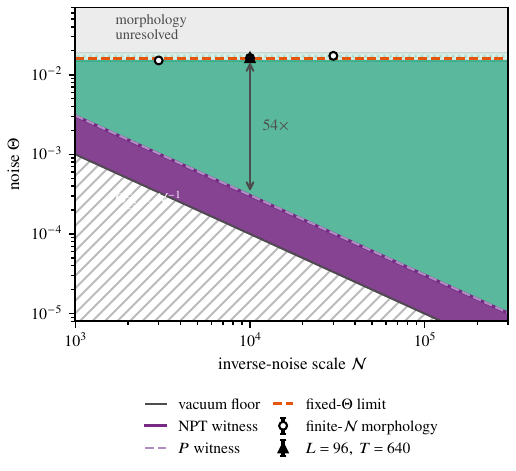}
 \caption{\textbf{Parametric separation of Gaussian witness and stripe nematic scales.} The NPT and second-moment $P$ thresholds on the exact $k_*$ shell decrease as $\cN^{-1}$. Finite-$\cN$ morphology estimates approach the dashed limit obtained from the first-moment equation at fixed $\Theta$. The hatched region lies below the vacuum-noise floor. Between the Gaussian curves and the morphology scale, resolved stripes remain after both witness margins on the exact $k_*$ shell have crossed zero. The filled triangle gives the $L=96$, $T=640$ estimate. At $\cN=10^4$, the morphology threshold is about $54$ times the NPT threshold, one point on a separation that grows with $\cN$.}
 \label{fig:phase}
\end{figure}

The long-time ensemble gives $\Theta_M^c\simeq1.64\times10^{-2}$. Direct integration of the first-moment equation at fixed $\Theta$ gives $\Theta_M^{\rm lim}\simeq1.60\times10^{-2}$. Lattice size, time window, and morphology criterion remain fixed as $\cN\to\infty$. The Supplemental Material gives statistical intervals and sensitivity checks.

\section{Discussion}
First moments and canonical covariance depend on different noise variables. The interval with resolved stripes and negative shell margins therefore grows with $\cN$. The multimode covariance sector persists to larger bath occupation, and broken translations produce nonzero covariance asymmetry at zero bath occupation.

Together, spatial imaging or Bragg measurements and covariance reconstruction near $k_*$ distinguish macroscopic order from the Gaussian correlations examined here. Similar separation may occur in photonic arrays, polariton fluids, and other bosonic lattices in which first moments carry order and second moments carry the nonclassicality tested here.

\section*{Data Availability}
The codes for quantum Turing patterns are available in the author's GitHub repository \url{https://github.com/IKEDAKAZUKI/Quantum-Turing-Pattern}. Reproduction data for this study may be available upon reasonable request to the author.

\FloatBarrier
\renewcommand{\bibfont}{\footnotesize}
\setlength{\bibsep}{0pt plus 0.2pt}
\bibliographystyle{apsrev4-2}
\bibliography{references}

\clearpage
\onecolumngrid
\begin{center}
 {\large\bfseries Supplemental Material: A Zoology of Quantum Turing Patterns}
\end{center}


This Supplement gives the Lindblad model and quantities for morphology, momentum-space covariance, fixed-registry calculations, weak pinning, covariance feedback, and long-time statistics. Calculations compare two noise scales for the same ordering mode and quantify covariance asymmetry from broken translations.

\section{Model and stable stripe backgrounds}
\begin{figure}[t]
 \includegraphics[width=0.97\textwidth]{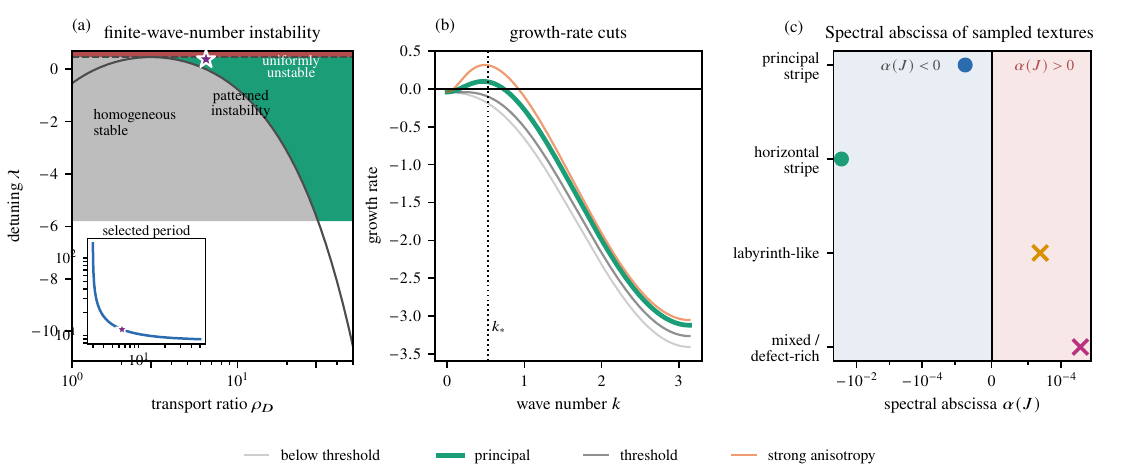}
 \caption{\textbf{Model and linear stability.} The finite-wave-number instability wedge appears in panel (a). The white-edged star marks $(\lambda,\rho_D)=(0.38,3+2\sqrt3)$, and the inset gives the same operating point in the period representation. It selects period 12 and corresponds to branch P in Main Fig.~1. Growth-rate cuts for $\rho_D=2$, $3$, $3+2\sqrt3$, and $15$ are compared in panel (b), with the dotted line at $k_*$. Panel (c) contrasts the spectral abscissae of the two stationary stripes with instantaneous Jacobians evaluated on two textures sampled at late times. Circles mark stationary solutions and crosses mark sampled textures.}
 \label{fig:S1}
\end{figure}

At each lattice site, let $a_{\bm r}$ annihilate a boson and define the scaled quadratures
\begin{equation}
 \hat q_{\bm r}=\frac{a_{\bm r}+a_{\bm r}^\dagger}{\sqrt{2\cN}},
 \qquad
 \hat p_{\bm r}=\frac{a_{\bm r}-a_{\bm r}^\dagger}{i\sqrt{2\cN}},
 \qquad
 [\hat q_{\bm r},\hat p_{\bm r'}]=\frac{i}{\cN}\delta_{\bm r\bm r'}.
 \label{eq:scaledqp}
\end{equation}
For numerical work the canonical vector is ordered as
\begin{equation}
 X=(q_1,\ldots,q_{L^2},p_1,\ldots,p_{L^2})^{\mathsf T}.
 \label{eq:canonicalOrderingS}
\end{equation}
The Lindblad generator is
\begin{equation}
 \dot\rho=-i[H,\rho]+\sum_\alpha\Diss[L_\alpha]\rho,
 \qquad
 \Diss[L]\rho=L\rho L^\dagger-\tfrac12\{L^\dagger L,\rho\}.
 \label{eq:masterS}
\end{equation}
Here $\rho$ is the density operator, $H$ is the Hamiltonian, and $L_\alpha$ are jump operators indexed by the dissipative channel $\alpha$. We write $H=H_{\rm loc}+H_\chi+H_{\rm bond}$. The cubic Hamiltonian is obtained by fully symmetrizing the noncommuting factors $\hat q_{\bm r}$ and $\hat p_{\bm r}$:
\begin{align}
 H_{\rm loc}&=\sum_{\bm r}\left[\omega a_{\bm r}^\dagger a_{\bm r}
 +\frac{i\epsilon}{2}\left(a_{\bm r}^{\dagger2}-a_{\bm r}^2\right)\right],
 \qquad \epsilon=2,\nonumber\\
 H_\chi&=\cN\chi_2\sum_{\bm r}\left[
 \frac{\hat q_{\bm r}^{2}\hat p_{\bm r}+\hat q_{\bm r}\hat p_{\bm r}\hat q_{\bm r}+\hat p_{\bm r}\hat q_{\bm r}^{2}}{3}
 -\frac{\hat p_{\bm r}^{3}}{3}\right].
 \label{eq:HlocalS}
\end{align}
The index $\mu\in\{x,y\}$ labels a lattice direction and $\hat\mu$ is the corresponding unit lattice vector. For $d_{\bm r\mu}=a_{\bm r}-a_{\bm r+\hat\mu}$, the bond channel is
\begin{align}
 L_{\bm r\mu}^{\rm bond}=\sqrt{2K_\mu}\,d_{\bm r\mu},\qquad
 H_{\rm bond}=\sum_{\bm r,\mu}\frac{iG_\mu}{2}
 \left(d_{\bm r\mu}^{\dagger2}-d_{\bm r\mu}^{2}\right).
 \label{eq:bondS}
\end{align}
It gives $D_{q\mu}=K_\mu-G_\mu$ and $D_{p\mu}=K_\mu+G_\mu$. We set $D_{qx}=1$, $D_{px}=\rho_D$, and $\eta_{\rm iso}=1+\delta\eta$. The atlas uses
\begin{equation}
 D_{qy}=(1-\eta_{\rm iso})D_\perp+\eta_{\rm iso}D_{qx},
 \qquad
 D_{py}=(1-\eta_{\rm iso})D_\perp+\eta_{\rm iso}D_{px},
 \qquad D_\perp=4.
 \label{eq:transportMapS}
\end{equation}
At $\delta\eta=0$ the interpolation is isotropic, and crossing this point reverses the preferred stripe direction. Parameters for the principal stripe are
\[
 (\lambda,\delta\eta,\chi_2,\nu)=(0.38,-0.28,-0.32,4),
 \qquad (D_{qx},D_{qy},D_{px},D_{py})=(1,1.84,6.46,5.77).
\]
For the horizontal stripe, the parameters are
\[
 (\lambda,\delta\eta,\chi_2,\nu)=(0.38,0.20,0.20,4),
 \qquad (D_{qx},D_{qy},D_{px},D_{py})=(1,0.40,6.46,6.96).
\]
Fig.~\ref{fig:S1}(a) shows the longitudinal $k_y=0$ stability slice, which depends only on $D_{qx}=1$ and $D_{px}=\rho_D$.

Local channels are
\begin{equation}
 L_{\bm r,-}=\sqrt{\kappa(n_{\rm th}+1)}a_{\bm r},
 \qquad
 L_{\bm r,+}=\sqrt{\kappa n_{\rm th}}a_{\bm r}^\dagger,
 \qquad
 L_{\bm r,2}=\sqrt{\gamma/\cN}\,a_{\bm r}^2,
 \label{eq:localjumpsS}
\end{equation}
with $\kappa=2$ and $\gamma=8$. Writing $\langle a_{\bm r}\rangle=\sqrt{\cN}(q_{\bm r}+ip_{\bm r})/\sqrt2$ gives the drift in the main text. The present quadrature normalization gives $\nu=\gamma/2=4$. The leading Wigner diffusion contains the local term $[\kappa(n_{\rm th}+1/2)+\gamma(q^2+p^2)]/\cN$ and the correlated bond increments \cite{GardinerZoller2004,VanKampen2007}. Let $F(X)$ denote the deterministic first-moment drift and let $X_*$ be the stationary stripe background. Define the canonical fluctuation $\xi=\sqrt{\cN}(X-X_*)$ and its symmetrized covariance by $V_{ij}=\langle\xi_i\xi_j+\xi_j\xi_i\rangle/2$. The linearized drift is $J=\partial_XF(X_*)$. These fluctuations obey
\begin{equation}
 \dot V=JV+VJ^{\mathsf T}+D_0+n_{\rm th}D_T,
 \qquad D_T=\kappa I_{\rm can},
 \label{eq:LyapunovS}
\end{equation}
where $I_{\rm can}$ is the identity on the full canonical covariance space. Write $q_{*,\bm r}$ and $p_{*,\bm r}$ for the sitewise quadrature components of the stationary stripe background $X_*$. With $K_\mu=(D_{q\mu}+D_{p\mu})/2$, the diffusion matrix at zero temperature is
\begin{equation}
 D_0=\operatorname{diag}(\Lambda_0,\Lambda_0),
 \qquad
 \Lambda_0=\operatorname{diag}_{\bm r}\!\left[\frac{\kappa}{2}+\gamma(q_{*,\bm r}^{2}+p_{*,\bm r}^{2})\right]+K_x(-\Delta_x)+K_y(-\Delta_y).
 \label{eq:D0S}
\end{equation}
Periodic boundary conditions give $(\Delta_\mu f)_{\bm r}=f_{\bm r+\hat\mu}+f_{\bm r-\hat\mu}-2f_{\bm r}.$

Scaled first moments have local thermal diffusion coefficient
\begin{equation}
 \frac{\kappa(n_{\rm th}+1/2)}{\cN}
 =\frac{\kappa}{2}\Theta,
 \qquad \Theta=\frac{2n_{\rm th}+1}{\cN}.
 \label{eq:noiseScalingS}
\end{equation}
Canonical covariance contains the additive term $n_{\rm th}D_T$. We call a background Hurwitz stable when its spectral abscissa $\alpha(J)=\max\operatorname{Re}\sigma(J)$ is negative, where $\sigma(J)$ is the spectrum of $J$. For a simple witness zero with smooth covariance feedback,
\begin{equation}
 n_{G,k_*}^{c}(\cN)=n_{G,k_*}^{c,(0)}+O(\cN^{-1}),
 \qquad
 \Theta_{G,k_*}^{c}(\cN)=\frac{2n_{G,k_*}^{c,(0)}+1}{\cN}+O(\cN^{-2}).
 \label{eq:leadingThresholdS}
\end{equation}
Here the superscript $(0)$ denotes the Gaussian calculation before covariance feedback. A simple zero satisfies $\partial w_G/\partial n_{\rm th}\ne0$ at $n_{G,k_*}^{c,(0)}$. The numerical slopes are nonzero at both crossings. At fixed $\Theta$ and $\cN\to\infty$, $n_{\rm th}=(\cN\Theta-1)/2$ and the first-moment equation retains the one-photon diffusion $\kappa\Theta/2$. The state-dependent two-photon and bond noise terms vanish as $\cN^{-1}$. Section~\ref{sec:longtimeS} integrates this limiting equation directly.

For the longitudinal slice $D_{qx}=1$ and $D_{px}=\rho_D$, the finite-wave-number wedge is
\begin{equation}
 \rho_D>3,
 \qquad
 2\sqrt3-3-\frac{(\rho_D-3)^2}{4\rho_D}<\lambda<2\sqrt3-3.
 \label{eq:wedgeS}
\end{equation}
Choosing $\rho_D=3+2\sqrt3$ selects $k_*=\pi/6$ and period 12. Both period-12 stripes are Hurwitz stable without registry pinning. The sampled nonstripe fields remain weakly time dependent and enter only instantaneous Jacobian comparisons.

\FloatBarrier

\section{Morphology checks}
For $R=q-\beta p$, with $\beta=(\sqrt3-1)/\sqrt{2\sqrt3}$, define
\begin{equation}
 \bar R=L^{-2}\sum_{\bm r}R_{\bm r},
 \qquad
 R_{\bm k}=L^{-2}\sum_{\bm r}(R_{\bm r}-\bar R)e^{-i\bm k\cdot\bm r},
 \qquad
 S(\bm k)=|R_{\bm k}|^2.
 \label{eq:fourierConventionS}
\end{equation}
Set $\theta_{\bm k}=\arg(k_x+ik_y)$ and use
\begin{equation}
 \mathcal S=\{\bm k\mathrel{|}\bigl||\bm k|-k_*\bigr|\le0.14\}.
 \label{eq:shellDefS}
\end{equation}
All momenta in this section use unnormalized lattice units. The stripe nematic moment, shell concentration, and real-space contrast are
\begin{align}
 \PhiS=\frac{\sum_{\bm k\in\mathcal S}S(\bm k)e^{2i\theta_{\bm k}}}{\sum_{\bm k\in\mathcal S}S(\bm k)},
 \qquad C_{\rm shell}=\frac{\sum_{\bm k\in\mathcal S}S(\bm k)}{\sum_{\bm k\ne0}S(\bm k)},\qquad
 \sigma_R=\left[L^{-2}\sum_{\bm r}(R_{\bm r}-\bar R)^2\right]^{1/2}.
 \label{eq:morphS}
\end{align}
Writing $f_+=L^{-2}\sum_{\bm r}\mathbf 1[R_{\bm r}>\bar R]$, where $\mathbf 1[\cdot]$ denotes the indicator function, the signed area imbalance is $f_+-1/2$. The normalized shell entropy is
\begin{equation}
 H_{\rm shell}=-\frac{\sum_{\bm k\in\mathcal S}p_{\bm k}\ln p_{\bm k}}{\ln N_{\mathcal S}},
 \qquad
 p_{\bm k}=\frac{S(\bm k)}{\sum_{\bm q\in\mathcal S}S(\bm q)},
 \qquad
 N_{\mathcal S}=|\mathcal S|.
 \label{eq:entropyS}
\end{equation}
We set $0\ln0=0$. When the normalization vanishes numerically, the normalized quantity is set to zero. Static atlas cells are displayed only when $\sigma_R\ge0.015$ and $C_{\rm shell}\ge0.10$.

For the Fourier maps in Main Fig.~3,
\begin{equation}
 S_{\rm exc}(\bm k)=\max\{S(\bm k)-S_{\rm bg},0\},
 \qquad
 S_{\rm bg}=\operatorname*{median}_{\bm k\in\mathcal B}S(\bm k),
 \label{eq:SexcS}
\end{equation}
with $\mathcal B=\{\bm k\mathrel{|}0.2<|\bm k|<1.2,\ \bigl||\bm k|-k_*\bigr|>0.14\}$. The spectral background is evaluated separately for each snapshot. Define
\begin{equation}
 d(\bm k)=\min\{|\bm k-(k_*,0)|,|\bm k+(k_*,0)|\},
 \qquad
 \mathcal P=\{\bm k\mathrel{|}d(\bm k)\le0.36\}.
 \label{eq:peakNeighborhoodS}
\end{equation}
We quantify peak broadening and concentration by
\begin{equation}
 \Delta k_{\rm peak}=\left[\frac{\sum_{\mathcal P}S_{\rm exc}(\bm k)d(\bm k)^2}{\sum_{\mathcal P}S_{\rm exc}(\bm k)}\right]^{1/2},
 \qquad
 C_{\rm Bragg}=\frac{\sum_{d(\bm k)=0}S_{\rm exc}(\bm k)}{\sum_{\mathcal P}S_{\rm exc}(\bm k)}.
 \label{eq:braggDiagnosticsS}
\end{equation}
A numerically vanishing peak normalization gives $\Delta k_{\rm peak}=C_{\rm Bragg}=0$. Let $\mathcal K_{\rm disp}=\{\bm k:|k_x|\le2k_*,\ |k_y|\le2k_*\}$ denote the momentum window displayed in Main Fig.~3. Its four logarithmic maps use
\begin{equation}
 S_{\rm exc,max}=\max_{j=1,\ldots,4}\ \max_{\bm k\in\mathcal K_{\rm disp}}S_{{\rm exc},j}(\bm k),
 \label{eq:SexcMaxS}
\end{equation}
where $j$ labels the four displayed snapshots.

Thermal trajectories are generated with the first-order implicit-explicit Euler--Maruyama scheme. The nonlinear drift and stochastic increments are explicit, while the linear reaction and transport operator is solved implicitly in Fourier space. All crossover calculations use $\Delta t=0.1$, and observables are stored every unit of time. Time is measured in units of the linear $q$ growth rate.

The $L=96$ ensemble uses $\cN=10^4$, $T=640$, and the averaging window $320<t\le640$. Its noise grid is $\Theta=0.014$, $0.015$, $0.016$, $0.018$, and $0.020$, with six paired counter-based Philox pseudorandom streams. The limiting equation uses $L=48$, $T=320$, and $160<t\le320$ at nine noise values from $0.014$ to $0.018$, with six independent streams at each value. Horizontal-branch controls use six streams per noise value at $L=48$ and $T=320$.

\begin{figure}[t]
 \includegraphics[width=0.97\textwidth]{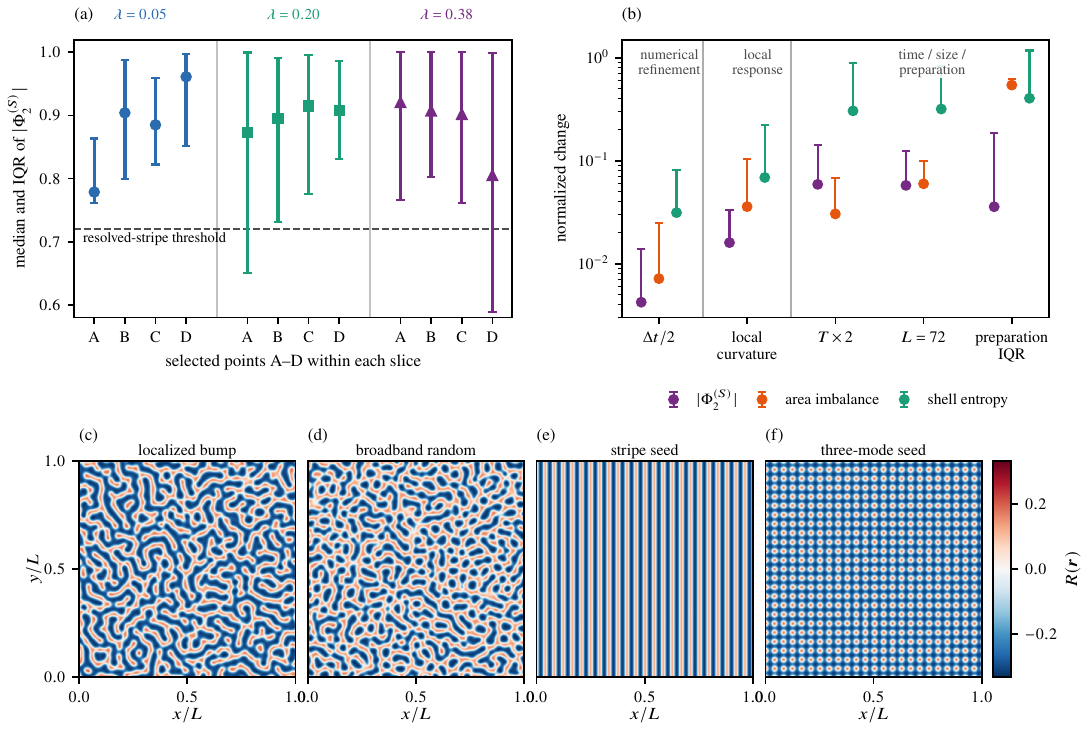}
 \caption{\textbf{Dependence on preparation and numerical settings.} Panel (a) shows the median and interquartile range from eight preparations at four selected points in each $\lambda$ slice. Labels A through D identify the points within a slice. The dashed line marks $|\PhiS|=0.72$. Panel (b) shows eight high-curvature sensitivity calculations normalized by the 5 to 95 percent range of each observable. Vertical separators distinguish numerical refinement, local parameter response, and sensitivity to time, size, and preparation. Markers show medians. Upper whiskers show maxima. Panels (c) through (f) show fields at late times from localized bump, broadband random, stripe, and three mode preparations at the same parameter point.}
 \label{fig:S2}
\end{figure}

Every atlas cell starts from the same deterministic preparation. The atlas is a continuation map from one preparation. Preparation dependence is examined separately. Subsampling the same arrays changes the extracted contour by less than two percent and changes fewer than one percent of the resolved cells. This is a resolution test on the same data. Fig.~\ref{fig:S2}(a) shows the spread over eight preparations. Within each $\lambda$ slice, resolved cells are ranked by $||\PhiS|-0.72|$. The 80 closest cells are sorted by $\chi_2$. Points A through D are four evenly spaced ranks in that list. Their coordinates are stored in the accompanying machine-readable file. The four Main text examples were selected to span visually distinct textures.

Panel (b) uses eight sensitivity points selected from the largest normalized second differences of area imbalance or shell entropy, subject to a minimum separation on the parameter grid. These points probe locally sharp regions of the atlas. Their coordinates and selection order accompany the data. Panels (c) through (f) show four preparations at the point used for the preparation control.

\FloatBarrier

\section{Gaussian covariance across momentum}
For a tested pair covariance $V_{\rm pair}$, $\widetilde\nu_-^{\rm PT}<1/2$ satisfies the negative partial transpose (NPT) criterion. With $I_4$ the identity on the pair covariance space, the condition $\lambda_{\min}(V_{\rm pair}-I_4/2)<0$ detects Gaussian nonclassicality in the Glauber--Sudarshan $P$ sense at the covariance level \cite{Simon2000,Weedbrook2012}. This $P$ witness is unrelated to the principal branch label P in Main Fig.~1. Exact $k_*$ directions satisfy
\begin{equation}
 k_x=\frac{2\pi n_x}{L},
 \qquad k_y=\frac{2\pi n_y}{L},
 \qquad n_x^2+n_y^2=(L/12)^2.
 \label{eq:exactShellS}
\end{equation}
On the tested commensurate lattices $L=60$, $120$, $180$, $240$, and $300$, the leading crossings on the exact $k_*$ shell are unchanged to numerical precision. Their values are
\begin{equation}
 n_{{\rm NPT},k_*}^{c}\simeq1.02,
 \qquad
 n_{P,k_*}^{c}\simeq1.04.
 \label{eq:shellThresholdsS}
\end{equation}
Let the 12 pairs $(q_j,p_j)$ label the sites of one period-12 stripe cell and set
\begin{equation}
 A=\left[\frac1{12}\sum_{j=0}^{11}(q_j^2+p_j^2)\right]^{1/2},
 \qquad
 \phi_m=\frac{2\pi m}{64}.
 \label{eq:referenceAmplitudeS}
\end{equation}
For each phase, define the spatially uniform field
\begin{equation}
 q_{\bm r,\phi_m}^{(0)}=A\cos\phi_m,
 \qquad
 p_{\bm r,\phi_m}^{(0)}=A\sin\phi_m
 \quad\text{for every site }\bm r.
 \label{eq:referenceFieldS}
\end{equation}
Let $X_{\phi_m}^{(0)}$ denote the resulting $2L^2$-component canonical vector. Constant full-lattice drives
\begin{equation}
 f_{\rm hold}^{(\phi_m)}=-F[X_{\phi_m}^{(0)}]
 \label{eq:holdingForceS}
\end{equation}
keep the 64 amplitude-matched states stationary without changing their Jacobians or diffusion matrices. Among these states, 38 have Hurwitz-stable Jacobians and define the uniform comparison family used below. We use
\begin{equation}
 Z_{\rm ref}(\bm k)=\frac{n_{{\rm NPT},\,\rm stripe}^{c}(\bm k)-\mu_{\rm ref}(\bm k)}{s_{\rm ref}(\bm k)}
 \label{eq:ZrefS}
\end{equation}
as a descriptive standardization over this Hurwitz-stable phase subset. Here $\mu_{\rm ref}$ and $s_{\rm ref}$ are its mean and sample standard deviation. Points with numerically vanishing spread are omitted from the standardization. Main Fig.~2(b) displays $Z_{\rm ref}$ on a linear scale from $-3$ to $0.25$, with zero mapped to white. Values below $-3$ are clipped to the lower endpoint. The raw minimum at $2k_*$ lies below the displayed range, while the largest finite value is about $0.19$. This family reproduces most of the broad pair profile. The stripe contrast is localized near $2k_*$. The ultraviolet endpoint approaches the Brillouin zone $M$ point, which is invariant under momentum inversion.

Each period-12 Bloch sector contains the reciprocal harmonics $k_x+m k_*$ for $m=0,\ldots,11$. Transverse sectors invariant under momentum inversion use the corresponding division between negative and nonnegative $k_x$.

Main Fig.~2(c) uses the fixed-registry $L=60$ sector with $|n_y|=29$, $|k_y|=29\pi/30$, and reduced residue $n_x\bmod5=0$. The two transverse signs give 24 traveling modes. Among the $L=60$ sectors included in the scan, this sector has the largest tested half-plane NPT crossing. Its fixed bipartition transposes the negative-$k_y$ modes against the positive-$k_y$ modes.

We evaluate
\begin{equation}
 E_N^{\rm full}=\sum_\alpha\max\{0,-\log_2(2\widetilde\nu_\alpha^{\rm PT})\},
 \qquad
 E_N^{\rm pair}=\sum_{\{\bm k,-\bm k\}}E_N[V_{\bm k,-\bm k}].
 \label{eq:multimodeS}
\end{equation}
The index $\alpha$ runs over the symplectic eigenvalues of the partially transposed Bloch-sector covariance. For any pair covariance $W$, $E_N[W]$ denotes the logarithmic negativity computed from its partially transposed symplectic spectrum. The second expression sums the results for separately reduced $4\times4$ pair covariances. Main Fig.~2(c) shades $E_N^{\rm full}-E_N^{\rm pair}$, and the full-sector result remains nonzero beyond the pairwise shell crossings.

\begin{figure}[t]
 \includegraphics[width=0.97\textwidth]{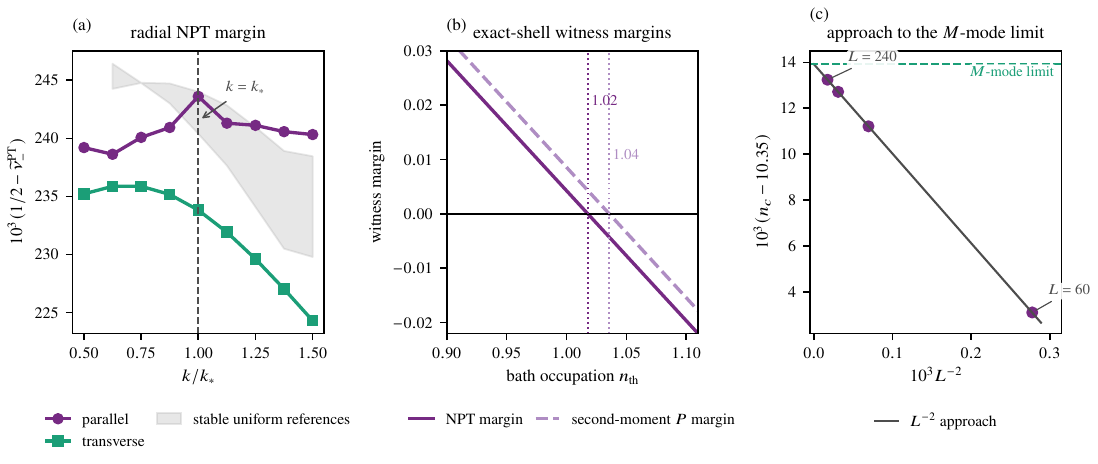}
 \caption{\textbf{Exact $k_*$ and ultraviolet Gaussian structure.} Radial NPT margins parallel and transverse to the stripe are shown in panel (a). The gray envelope spans the uniform comparison states and the arrow identifies the exact $k_*$ shell. NPT and second-moment $P$ margins on the exact $k_*$ shell appear in panel (b). Panel (c) follows the ultraviolet pair toward the limit at the Brillouin-zone $M$ point as the lattice size increases. Main Fig.~2(c) gives the full Bloch sector and pair comparison.}
 \label{fig:S3}
\end{figure}
\FloatBarrier

\section{Covariance at fixed registry}
\begin{figure}[t]
 \includegraphics[width=0.97\textwidth]{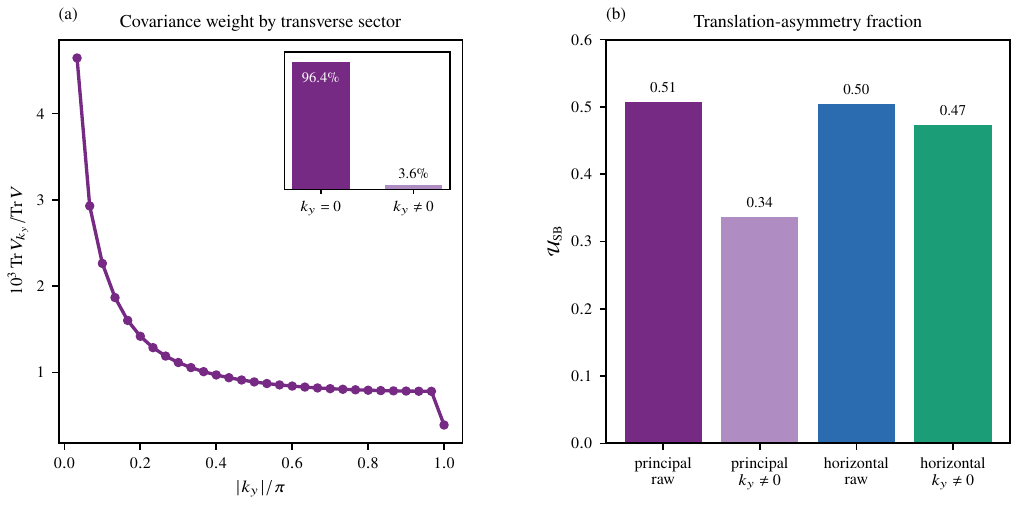}
 \caption{\textbf{Covariance at fixed registry.} Panel (a) resolves the covariance weight of the principal stripe across nonzero transverse sectors. The inset compares the $k_y=0$ contribution with the total from $k_y\ne0$. Translation asymmetry for both stripes is displayed in panel (b) before and after removal of the $k_y=0$ sector containing the nearly neutral translation mode. Averaging over registries changes the covariance witnesses on the exact $k_*$ shell only weakly.}
 \label{fig:S4}
\end{figure}

With $C=V-I_{\rm can}/2$ and $V_0=V|_{n_{\rm th}=0}$, let $T_s$ denote the orthogonal symplectic permutation that translates the canonical vector by $s$ lattice sites along the stripe direction. It acts along $x$ for the principal stripe. For the horizontal branch, we exchange the physical axes and work in coordinates aligned with the stripe, where translation has the same representation. Averaging the fluctuation covariance over translated registries gives
\begin{equation}
 \mathcal T_{\rm fluc}(C)=\frac1{12}\sum_{s=0}^{11}T_sCT_s^{\mathsf T}.
 \label{eq:TflucS}
\end{equation}
We define the normalized translation asymmetry as
\begin{equation}
 \USB=\frac{\|C-\mathcal T_{\rm fluc}(C)\|_F^2}{\|C\|_F^2}.
 \label{eq:USBS}
\end{equation}
In stripe-aligned coordinates, let $P_\perp$ project onto the canonical $k_y\ne0$ sector and set $C_\perp=P_\perp CP_\perp^{\mathsf T}$. The corresponding sector-excluded quantity is
\begin{equation}
 \USB^\perp=\frac{\|C_\perp-\mathcal T_{\rm fluc}(C_\perp)\|_F^2}{\|C_\perp\|_F^2}.
 \label{eq:USBperpS}
\end{equation}
Here $\|A\|_F=[\operatorname{Tr}(A^{\mathsf T}A)]^{1/2}$ is the Frobenius norm. These ratios quantify covariance asymmetry in the canonical basis. At zero bath occupation, the $k_y=0$ blocks of the principal stripe account for about $96\%$ of $\operatorname{Tr}V_0$. In the transverse-sector weight plotted in Fig.~\ref{fig:S4}(a), both members of each pair exchanged by momentum inversion are included. Removing the complete translation sector leaves nonzero asymmetry for both stripes. The uniform comparison states satisfy $\mathcal T_{\rm fluc}(C)=C$.

\FloatBarrier

\section{Weak pinning and the second harmonic pair}
The stripe registry is the spatial phase of its period-12 profile. We fix it with
\begin{equation}
 H_{\rm pin}=-\cN h_{\rm pin}\sum_{x,y}\cos(k_*x+\phi_*)
 \left(\hat q_{x,y}-\beta\hat p_{x,y}\right),
 \qquad \phi_*=\pi/12.
 \label{eq:pinS}
\end{equation}
Let $X(h_{\rm pin})$ be the aligned pinned background and let $A_\ell(h_{\rm pin})$ be the magnitude of its $\ell$th Fourier harmonic. With $X_0=X(h_{\rm pin}=0)$,
\begin{equation}
 \epsilon_{\rm pin}=\frac{\|X(h_{\rm pin})-X_0\|_2}{\|X_0\|_2},
 \qquad
 \delta A_\ell=\frac{A_\ell(h_{\rm pin})-A_\ell(0)}{A_\ell(0)}.
 \label{eq:pinDeformationS}
\end{equation}
Here $\|\cdot\|_2$ is the Euclidean norm on the first-moment vector. At the one percent boundary,
\begin{equation}
 \max\{\epsilon_{\rm pin},|\delta A_1|,|\delta A_2|\}=0.01,
 \qquad h_{\rm pin}^{1\%}\simeq2.3\times10^{-4}.
 \label{eq:pinCriterionS}
\end{equation}
Write the nonlinear fields as
\begin{equation}
 F_\chi(q,p)=\begin{pmatrix}\chi_2(q^2-p^2)\\-2\chi_2qp\end{pmatrix},
 \qquad
 F_\gamma(q,p)=-\nu(q^2+p^2)\begin{pmatrix}q\\p\end{pmatrix},
 \qquad F_{\rm nl}=F_\chi+F_\gamma.
 \label{eq:nonlinearFieldsS}
\end{equation}
For a vector field $F$,
\begin{equation}
 [\nabla^2F\mathbin{:}V]_i=\sum_{a,b}\frac{\partial^2F_i}{\partial X_a\partial X_b}V_{ab}.
 \label{eq:hessianContractionS}
\end{equation}
The ordering correction is determined by
\begin{equation}
 \Delta F_{\rm ord}[V]
 =\frac12\nabla^2F_\chi\mathbin{:}V
 +\frac12\nabla^2F_\gamma\mathbin{:}(V-I_{\rm can}/2).
 \label{eq:orderingCorrectionS}
\end{equation}
Equation~\eqref{eq:backgroundResponseS} defines the background correction $\delta X$. We monitor the pinned covariance through
\begin{equation}
 \epsilon_{\rm cov}=\frac{\operatorname{Tr}V}{\cN\|X\|_2^2},
 \qquad
 \epsilon_F=\frac{\|\Delta F_{\rm ord}\|_2}{\cN\|F_{\rm nl}(X)\|_2},
 \qquad
 \epsilon_X=\frac{\|\delta X\|_2}{\|X\|_2}.
 \label{eq:pinResponseRatiosS}
\end{equation}
All three remain below $3\%$ at the boundary, and the linearized drift remains Hurwitz stable. Weak pinning preserves the sign of the $2k_*$ response. The detuning-tuned control keeps the principal transport and nonlinear parameters. It sets $\lambda=-0.10$ and matches $|A_1(h_{\rm pin})-A_1(0)|$ to the stripe response. The transport-tuned control keeps the principal detuning and nonlinear parameters. It sets $D_{px}=2.5$ and matches
\begin{equation}
 R_{\rm glob}(h_{\rm pin})=\|X(h_{\rm pin})-X_0\|_2
 \label{eq:globalResponseNormS}
\end{equation}
to the stripe response. Let an overbar denote the mean over the 38 unpinned uniform comparison states. Panels (a) and (d) use
\begin{align}
 \Delta_{\rm held}\widetilde\nu_-^{\rm PT}(h)
 &=\widetilde\nu_{-,\rm stripe}^{\rm PT}(h)
 -\overline{\widetilde\nu_{-,\rm held}^{\rm PT}}(0),\nonumber\\
 \Delta_{\rm held}n_{\rm NPT}^{c}(h)
 &=n_{\rm NPT,stripe}^{c}(h)
 -\overline{n_{\rm NPT,held}^{c}}(0).
 \label{eq:heldDifferencesS}
\end{align}
The controls in panel (c) are compared at pinning strengths chosen to match the induced fundamental amplitude,
\begin{equation}
 \Delta_{\rm ctrl}n_{\rm NPT}^{c}
 =n_{\rm NPT,stripe}^{c}(h_{\rm stripe})
 -n_{\rm NPT,ctrl}^{c}(h_{\rm ctrl}).
 \label{eq:controlDifferenceS}
\end{equation}
For the Bloch-reduced opposite-momentum pair, let $a_+=a_{(2k_*,0)}$ and $a_-=a_{(-2k_*,0)}$. Define $n_\pm=\langle a_\pm^\dagger a_\pm\rangle$ and $m=\langle a_+a_-\rangle$. With $I_2$ the identity on a single-mode quadrature space, define
\begin{equation}
 V_{\rm sur}=\begin{pmatrix}
 (n_++\tfrac12)I_2&B(m)\\ B(m)^{\mathsf T}&(n_-+\tfrac12)I_2
 \end{pmatrix},
 \qquad
 B(m)=\begin{pmatrix}\operatorname{Re}m&\operatorname{Im}m\\\operatorname{Im}m&-\operatorname{Re}m\end{pmatrix}.
 \label{eq:pairingSurrogateS}
\end{equation}
Retaining $n_\pm$ and $m$, the surrogate sets normal intermode coherence, local quadrature anisotropy, and the remaining cross correlations to zero. Let $\Delta_{\rm held}\widetilde\nu_{-,\rm sur}^{\rm PT}$ denote its shift relative to the same held-family mean and define the remainder
\begin{equation}
 \Delta_{\rm rem}
 =\Delta_{\rm held}\widetilde\nu_{-,\rm full}^{\rm PT}
 -\Delta_{\rm held}\widetilde\nu_{-,\rm sur}^{\rm PT}.
 \label{eq:surrogateRemainderS}
\end{equation}
The surrogate reproduces the sign and overestimates the full-covariance shift, whose value depends on the comparison state.

\begin{figure}[t]
 \includegraphics[width=0.97\textwidth]{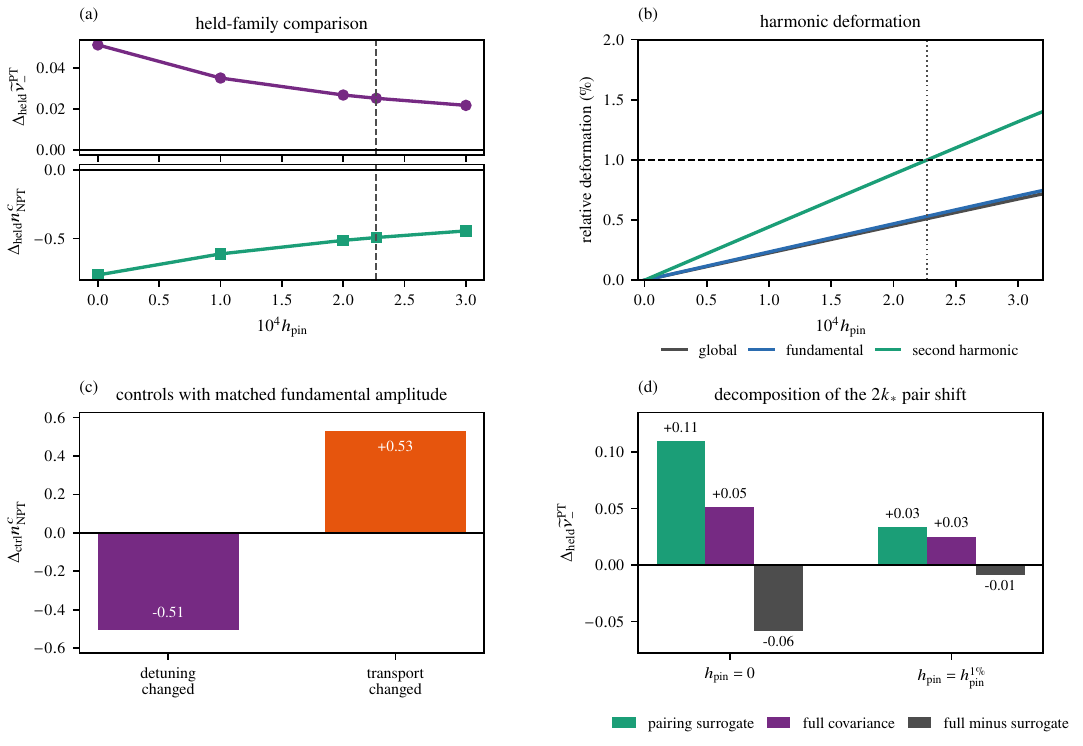}
 \caption{\textbf{Weak pinning response of the $2k_*$ pair.} Panel (a) shows $\Delta_{\rm held}\widetilde\nu_-^{\rm PT}$ and $\Delta_{\rm held}n_{\rm NPT}^{c}$ relative to the mean of the unpinned uniform comparison family. Background deformation and the first two harmonics appear in panel (b), whose maximum defines $h_{\rm pin}^{1\%}$. Panel (c) shows $\Delta_{\rm ctrl}n_{\rm NPT}^{c}$ for controls obtained by changing the detuning or transport while matching the induced fundamental amplitude. Panel (d) decomposes the held-family partial transpose shift into the occupation-pairing surrogate, the full covariance, and their difference $\Delta_{\rm rem}$.}
 \label{fig:S5}
\end{figure}

\section{Covariance iteration with feedback}
Let $X^{(0)}$ be the pinned first-moment background before covariance feedback, and let $X^{(j)}$ denote its $j$th update. For trial bath occupation $n$ and correction $\mathcal K^{(j-1)}$, the map $\mathcal L^{-1}$ returns the stationary covariance from
\begin{equation}
 [J(X^{(j)})+\mathcal K^{(j-1)}]V+V[J(X^{(j)})+\mathcal K^{(j-1)}]^{\mathsf T}
 +D_0(X^{(j)})+nD_T=0.
 \label{eq:LyapIterationS}
\end{equation}
Differentiation in $\partial_X\Delta F_{\rm ord}[V]$ is performed at fixed $V$. Here $t_x$ is the translation tangent, $\ell_x$ is the corresponding left mode normalized by $\ell_x^{\mathsf T}t_x=1$, and $\zeta$ enforces the phase condition. The covariance induced background correction solves
\begin{equation}
 \begin{pmatrix}J&t_x\\ \ell_x^{\mathsf T}&0\end{pmatrix}
 \begin{pmatrix}\delta X\\ \zeta\end{pmatrix}
 =-\begin{pmatrix}\cN^{-1}\Delta F_{\rm ord}[V]\\0\end{pmatrix},
 \qquad \ell_x^{\mathsf T}\delta X=0.
 \label{eq:backgroundResponseS}
\end{equation}
The coupled iteration reads
\begin{align}
\begin{aligned}
 V^{(j)}&=\mathcal L^{-1}[X^{(j)},n_c^{(j)},\mathcal K^{(j-1)}],\\
 \mathcal K^{(j)}&=\cN^{-1}\partial_X\Delta F_{\rm ord}[V^{(j)}]\big|_{X^{(j)}},\\
 X^{(j+1)}&=X^{(0)}+\delta X[V^{(j)}],\\
 n_c^{(j+1)}&=\mathcal C[X^{(j+1)},\mathcal K^{(j)}].
 \label{eq:iterationS}
 \end{aligned}
\end{align}
Here $n_c=n_{{\rm NPT},2k_*}^{c,{\rm sc}}$, where ${\rm sc}$ denotes the self-consistent covariance iteration. The functional $\mathcal C$ returns the numerical root of $w_{\rm NPT}=0$ for the $(2k_*,-2k_*)$ pair. We initialize $\mathcal K^{(-1)}=0$ and bracket each root before applying Brent's method. The root tolerance is $2\times10^{-8}$ and the iteration uses at most ten steps. For the covariance direct sum $\mathbb V^{(j)}=\bigoplus_s V_s^{(j)}$ with sector multiplicities $m_s$, define
\begin{align}
 r_n^{(j)}&=\frac{|n_c^{(j+1)}-n_c^{(j)}|}{\max(1,|n_c^{(j)}|)},\nonumber\\
 r_X^{(j)}&=\frac{\|X^{(j+1)}-X^{(j)}\|_2}{\|X^{(j)}\|_2},\nonumber\\
 r_V^{(j)}&=\frac{[\sum_s m_s\|V_s^{(j+1)}-V_s^{(j)}\|_F^2]^{1/2}}{[\sum_s m_s\|V_s^{(j)}\|_F^2]^{1/2}},\nonumber\\
 r_{\mathcal K}^{(j)}&=\frac{\|\mathcal K^{(j+1)}-\mathcal K^{(j)}\|_F}{\max(1,\|\mathcal K^{(j)}\|_F)}.
 \label{eq:iterationResidualsS}
\end{align}
All final residuals are below $2\times10^{-6}$.

The holding force remains fixed at its value before covariance feedback, and the same map updates every uniform comparison state. Thirty-five of the 38 initially Hurwitz-stable phases remain in the Hurwitz domain, while three leave it. After convergence, each surviving uniform-state background is refined by Newton's method at fixed covariance contractions, followed by recomputation of its crossing and Jacobian correction. The stripe converges to $n_{{\rm NPT},2k_*}^{c,{\rm sc}}\simeq1.91$, and the remaining uniform comparison states span $2.15$ to $2.25$. Crossings on the exact $k_*$ shell move by less than $0.3\%$.

\begin{figure}[t]
 \includegraphics[width=0.97\textwidth]{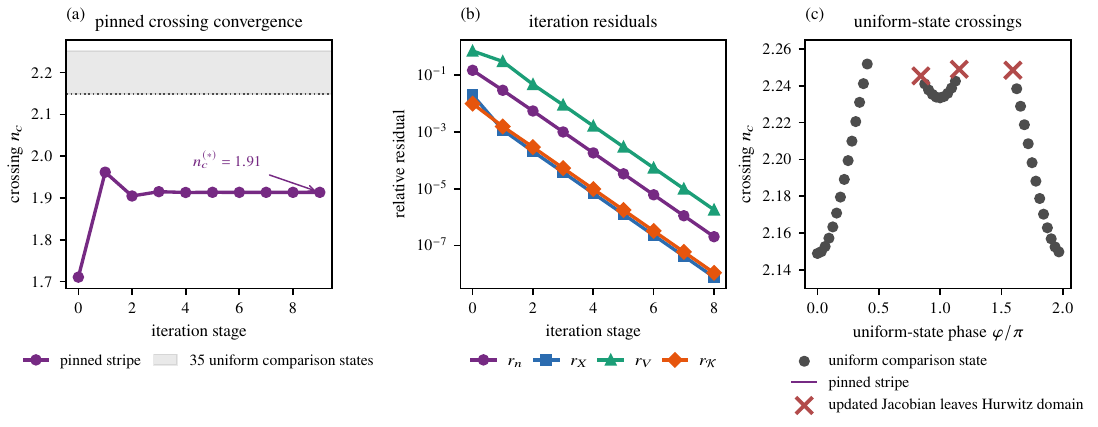}
 \caption{\textbf{Covariance iteration with feedback.} Panel (a) shows convergence of the stripe crossing relative to the range of the 35 Hurwitz-stable uniform comparison states. The range is shaded in pale gray. Panel (b) follows the crossing, background, covariance, and Jacobian-correction residuals. Panel (c) gives the phase-resolved crossings after the fixed-covariance Newton refinement defined in the text. Enlarged crosses mark the last iterates of three phases whose updated Jacobians leave the Hurwitz domain.}
 \label{fig:S6}
\end{figure}
\FloatBarrier

\section{Stripe nematic crossover at long times}\label{sec:longtimeS}
Thermal trajectories retain the local and bond noise terms of the same Lindblad family. The threshold estimator uses the trajectory ensemble, while Main Fig.~3 contains separate illustrative snapshots. Panels (a), (b), and (d) use stored arrays at $\Theta=10^{-4}$, $0.004$, and $0.512$. Their original stream and time identifiers were not retained. Panel (c) uses stream 0 at $t=230$ from the separate $L=192$, $T=320$ calculation and was selected by its minimum distance to the nematic cutoff. The available source and selection information accompanies the data. Each snapshot is centered and standardized separately, so color encodes morphology after removal of absolute contrast.

For trajectory $j$, define the late-time mean
\begin{equation}
 \bar\phi_j(\Theta)=\frac{1}{T_2-T_1}\int_{T_1}^{T_2}|\Phi_{2,j}^{(S)}(t)|\,dt,
 \qquad
 \bar\phi(\Theta)=\frac{1}{N_{\rm traj}}\sum_{j=1}^{N_{\rm traj}}\bar\phi_j(\Theta).
 \label{eq:morphEstimatorS}
\end{equation}
Here $T_1$ and $T_2$ are the endpoints of the late-time averaging window, and $N_{\rm traj}$ is the number of trajectories at that noise value.
The morphology crossover satisfies $\bar\phi(\Theta_M^c)=0.72$. The static atlas mask is not applied to these trajectories. Contrast in real space, Bragg concentration, and peak width are tracked separately.

Each trajectory is averaged over its late-time window before resampling. The $L=96$ calculation uses the same six Philox stream identities at every noise value and resamples them as pairs. Independent base seeds are used across noise for the $L=48$ and limiting calculations. We use the cutoff $|\PhiS|=0.72$ and repeat the calculation from $0.68$ to $0.76$. Each bootstrap curve is made decreasing by isotonic regression before linear interpolation. Raw linear, isotonic, and logistic estimates differ by less than $0.1\%$.

We report percentile bootstrap intervals from 5000 trajectory-level resamples. All $L=96$ resamples cross inside the simulated interval. Resampling the noise levels independently gives $[1.59,1.69]\times10^{-2}$, and leave-one-trajectory values span $[1.62,1.67]\times10^{-2}$. The limiting calculation also has 5000 in-range crossings, with leave-one-trajectory values from $[1.59,1.65]\times10^{-2}$. For the shorter $L=48$ calculation, 4297 of 5000 resamples cross inside the simulated interval.

For cutoffs from $0.68$ to $0.76$ and for the raw, isotonic, and logistic estimates, the morphology threshold remains above both Gaussian thresholds. At $\cN=10^4$, the morphology-to-NPT ratio is $54$. Holding the numerical NPT crossing fixed and propagating only the morphology bootstrap uncertainty gives a range of about $52$ to $56$.

\begin{figure}[t]
 \includegraphics[width=0.86\textwidth]{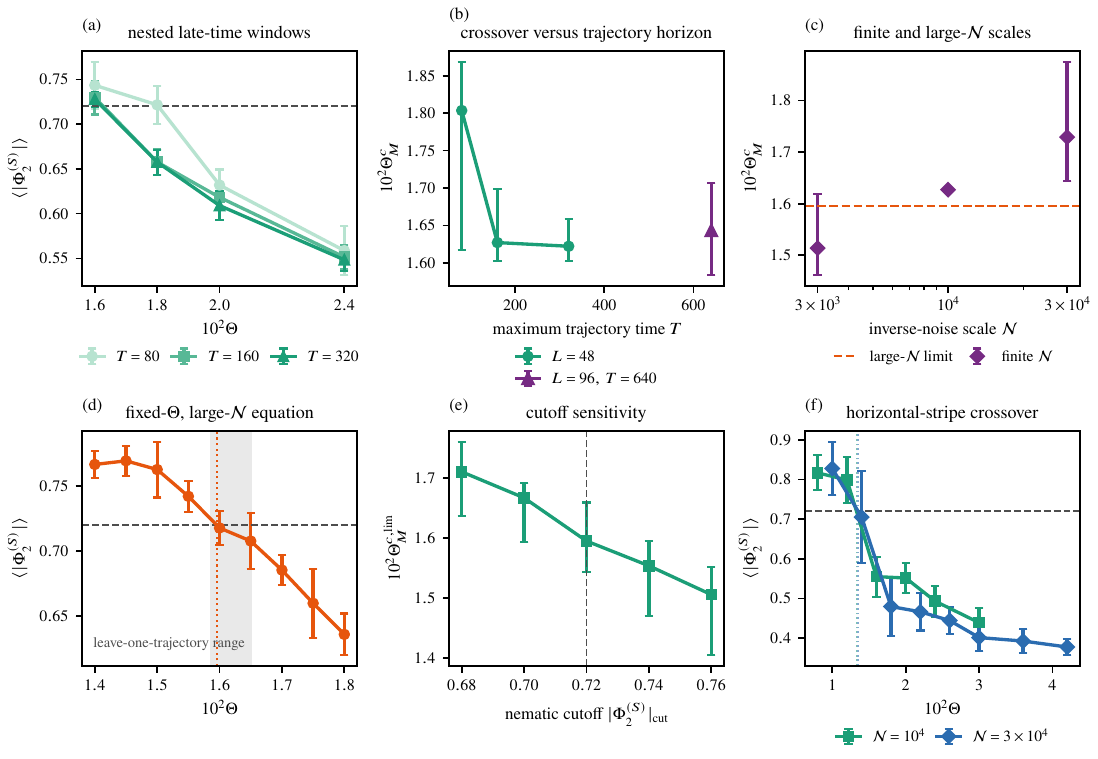}
 \caption{\textbf{Stripe nematic crossover versus time and $\cN$.} Panel (a) compares late-time windows. Panel (b) adds the separate $L=96$, $T=640$ estimate. Panels (c) and (d) compare finite-$\cN$ results with the direct large-$\cN$ equation, with the leave-one-trajectory range shaded in gray. Panels (e) and (f) vary the nematic cutoff and stripe branch. Noise axes use $10^2\Theta$.}
 \label{fig:S7}
\end{figure}
\FloatBarrier

The two morphology estimates are
\begin{align}
 \Theta_M^c&\simeq1.64\times10^{-2},&
 95\%\ \text{interval}&=[1.58,1.71]\times10^{-2},\nonumber\\
 \Theta_M^{\mathrm{lim}}&\simeq1.60\times10^{-2},&
 95\%\ \text{interval}&=[1.56,1.67]\times10^{-2}.
 \label{eq:thetaMS}
\end{align}
Here $\Theta_M^{\mathrm{lim}}$ is the $\cN\to\infty$ limit at fixed lattice size, time window, and morphology criterion. Cutoffs from $0.68$ to $0.76$ give $1.51\times10^{-2}$ to $1.71\times10^{-2}$, and the horizontal branch lies in the same range. Exact $k_*$ Gaussian crossings map to physical noise of order $\cN^{-1}$.

\renewcommand{\bibfont}{\scriptsize}
\setlength{\bibsep}{-0.5pt plus 0.2pt}
\bibliographystyle{apsrev4-2}

\end{document}